\documentclass{aa}

\usepackage{graphicx}
\usepackage{natbib}
\usepackage{color}
\usepackage{float}
\usepackage{txfonts}
\usepackage{placeins}
\usepackage{pdflscape}
\usepackage{mathtools, siunitx}
\usepackage{amsmath} 
\usepackage{amssymb} 
\usepackage{multicol} 
\usepackage{bm} 
\usepackage{epsfig}
\usepackage{multirow}
\usepackage{lscape}
\usepackage{subfig}
\usepackage[dvipsnames]{xcolor}

\begin{document}

   \title{First detailed study of two eccentric eclipsing binaries: \\ TYC 5378-1590-1 and TYC 8378-252-1 }

   \author{P. Zasche~\inst{1},
          D. S\"urgit~\inst{2,3},
          A. Erdem~\inst{2,4},
          C. A. Engelbrecht~\inst{5}, \and
          F. Marang~\inst{5}
          }
\offprints{Petr Zasche , \\
 \email{zasche@sirrah.troja.mff.cuni.cz}}

 \institute{
  $^{1}$ Astronomical Institute, Charles University, Faculty of Mathematics and Physics, CZ-180~00, Praha 8, V~Hole\v{s}ovi\v{c}k\'ach 2 Czech Republic \\
  $^{2}$ Astrophysics Research Center and Ulup{\i}nar Observatory, \c{C}anakkale Onsekiz Mart University, TR-17100, \c{C}anakkale, T\"{u}rkiye \\
  $^{3}$ Department of Space Sciences and Technologies, Faculty of Science, \c{C}anakkale Onsekiz Mart University, Terzio\u{g}lu Kamp\"{u}s\"{u}, TR-17100, \c{C}anakkale, T\"{u}rkiye \\
  $^{4}$ Department of Physics, Faculty of Science, \c{C}anakkale Onsekiz Mart University, Terzio\u{g}lu Kamp\"{u}s\"{u}, TR-17100, \c{C}anakkale, T\"{u}rkiye \\
  $^{5}$ Department of Physics, University of Johannesburg, PO Box 524, Auckland Park 2006, South Africa}

\titlerunning{Study of TYC 5378-1590-1 and TYC 8378-252-1}
\authorrunning{Zasche et al.}


\abstract{\textit{Aims}: The analysis of combined photometry and spectroscopy of eccentric eclipsing
binary systems facilitates the derivation of very precise values for a large ensemble of physical
parameters of the component stars and their orbits, thereby providing stringent tests of theories of
stellar structure and evolution. In this paper two eccentric eclipsing binary systems, TYC 5378-1590-1
and TYC 8378-252-1, are studied in detail for the first time.

\textit{Methods}: Radial velocities were obtained using cross-correlation methods applied to
mid-resolution spectra covering almost the entire orbital phase domains of these two systems. TESS
photometry was used for the analysis of TYC 5378-1590-1, whereas ASAS-SN photometry was used for
the analysis of TYC 8378-252-1.

\textit{Results}: We obtained the first precise derivation of the physical parameters of these systems.
Both systems display moderately eccentric orbits ($e \sim 0.3$ and $0.2$) with periods of 3.73235 and
2.87769 days, respectively. The apsidal motion is very slow, with a duration of several centuries for
both systems. We present two models for the apsidal motion of TYC 5378-1590-1. The internal structure
constant derived from observations for TYC 8378-252-1 is approximately $11\%$ lower than theoretical
predictions. We discuss possible reasons for this discrepancy. Our analysis indicates that the
components of both systems are on the main sequence. The components of TYC 5378-1590-1 are relatively
young stars (age $\sim 600$ Myr)  close to the ZAMS, whereas the components of TYC 8378-252-1 are
relatively old stars (age $\sim 4$ Gyr) close to the TAMS. Our finding that the circularization
timescale for TYC 5378-1590-1 is $\sim 200$ times longer than its evolutionary age is compatible with
circularization theory; however, our finding that the evolutionary age of TYC 8378-252-1 is approximately ten
times longer than the circulation age, while its orbital eccentricity is quite high ($e \sim 0.2$),
challenges the present theories of circularization. }


\maketitle


\section{Introduction}
\label{introduction}

The importance of eclipsing binaries for advancing our understanding of stellar structure and evolution
was reviewed by \cite{2012ocpd.conf...51S}. Eccentric eclipsing binaries contribute particular
insights in this respect \citep[see, e.g.,][]{2019A&A...628A..29C}. For eccentric systems,
high-precision photometry from TESS \citep{2015JATIS...1a4003R} enables tests of relativistic
predictions of apsidal motion to be performed with unprecedented accuracy \citep{2021A&A...649A..64B}.
A recent study \citep{2018ApJS..235...41K} implied a possible relation between the orbital period and
the apsidal motion period among known systems displaying apsidal motion; however, the possibility of
observational bias in this respect has not been discounted. Notwithstanding this caveat, the existence
of a hard upper limit to the eccentricity (which is a function of the orbital period) has been firmly
established in the observed period-eccentricity ($P-e$) diagram. The appearance of the observational
$P-e$ diagram corresponds to theoretical expectations based on the dynamics of the tidal
circularization process; however, tidal effects are expected to have diverse impacts on early- and
late-type stars because of their differing internal structures. A corresponding
divergence between the  behavior of eclipsing systems composed of hot stars and systems composed of
cool stars was demonstrated recently \citep{2016ApJ...824...15V}. These statements call for an
expansion of the number of eccentric systems for which physical parameters can be obtained with high
precision, so that theoretical models may be assessed more incisively. In this study, we present a
study of two previously unexplored eclipsing eccentric systems, using both photometry and spectroscopy.


\section{Sources of data}

The data for TYC 5378-1590-1 and TYC 8378-252-1 were obtained partly from dedicated observational
campaigns on these stars. This is specifically true of the spectroscopic data, where several dozen 
radial velocity observations were obtained during multi-year campaigns using the Cassegrain-mounted
spectrograph on the 1.9m telescope at the South African Astronomical Observatory (SAAO).


\subsection{Spectroscopy}
\label{spectroscopy}

\begin{figure*}
  \centering
    \includegraphics[scale=2.3]{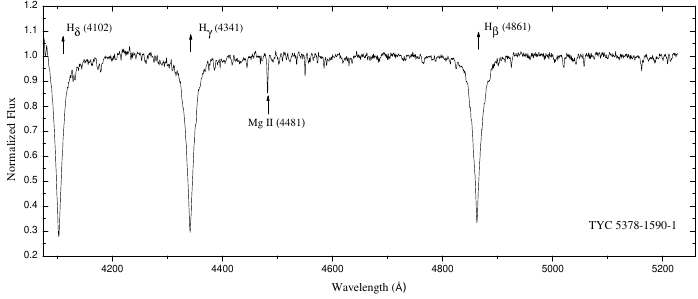} \\
    \includegraphics[scale=2.3]{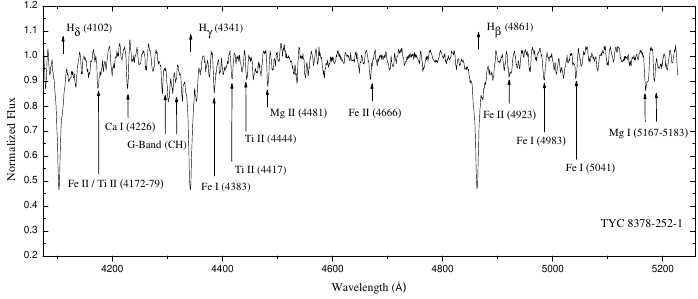}
\caption{Sample spectra for TYC 5378-1590-1 (upper diagram, for phase 0.56) and TYC 8378-252-1 (lower
diagram, for phase 0.01) obtained in this study. While the hydrogen Balmer lines and the Mg II (4481)
line are indicated in the spectrum of TYC 5378-1590-1, many metallic lines are also indicated in the
spectrum of TYC 8378-252-1. } \label{sample_spectra}
\end{figure*}

Spectroscopic observations of TYC 5378-1590-1 and TYC 8378-252-1 were conducted with the Spectrograph Upgrade: Newly Improved Cassegrain (SpUpNIC) instrument \citep[see][for
details]{Crause_etal_2016,Crause_etal_2019}  mounted at the Cassegrain focus of the 1.9m telescope at
the SAAO.

We selected grating 4 of the spectrograph. This 1200 lines/mm grating has a wavelength coverage of 400
-- 525 nm with a blaze peak at 510 nm and a resolution of 0.062 nm/px (corresponding to a resolution of
approximately 80 km s$^{-1}$ in radial velocity and an approximate resolving power $R = 3000$ at the
wavelength of the H$\beta$ line).  The wavelength range between 400 and 525 nm provided a spectral
interval that allowed an accurate derivation of radial velocities (RVs) of both components on most of
the spectra obtained. We used a slit width of 1.35 arcsec for all spectroscopic observations of the two
eclipsing systems and the associated RV standards, which were observed contemporaneously. A total of 62
and 86 spectra were obtained for TYC 5378-1590-1 and TYC 8378-252-1, respectively, across the 2019,
2020, and 2021 observing seasons.

HR 3383 (A1V, $V_r = 2.80$ km ${\rm s}^{-1}$) and HR 6031 (A1V, $V_r = -5.10$ km ${\rm s}^{-1}$) were
observed contemporaneously as standard stars for the RV measurements of the two binary systems. The
exposure times were set at the maximum feasible value of 1200 seconds for each spectrum because of the
faintness of the two targets (exposure times longer than 1200 s are usually compromised by a high
frequency of cosmic ray events in the spectra at the SAAO site). The relatively large errors in the
measured RV values (see Figs. \ref{FigRV_TYC5378} and \ref{FigRV_TYC8378}) are predominantly the result
of this upper limit on the feasible exposure time for these two relatively faint targets. Arc spectra
using a Cu-Ar lamp were taken as comparison spectra before and after each stellar spectrum. A set of
quartz-iodine lamp images was also taken every night for flat-field calibrations. Standard IRAF
procedures were used for the spectral data reduction and calibrations.

An example of the observed spectra for each of TYC 5378-1590-1 and TYC 8378-252-1 is shown in Fig.
\ref{sample_spectra}. The sample spectra in Fig. \ref{sample_spectra} were selected from the
conjunction phases of the components, with the more massive component transiting in front of the less
massive component in each instance. The characteristic lines of early A-type stars are seen in the
spectrum of TYC 5378-1590-1, whereas the characteristic lines of middle F-type stars are seen in the
spectrum of TYC 8378-252-1.


\subsection{Radial velocities}
\label{RVs}

The cross-correlation method was applied to derive RVs of the component stars of these two eccentric
binaries, using the IRAF package {\sc Fxcor} \citep{Tonry_Davis_1979,Popper_Jeong_1994}. The spectral
region around the Mg II (4481) line, which is the most prominent line apart from the hydrogen Balmer
lines (see Fig. \ref{sample_spectra}), was used for the RV calculations of TYC 5378-1590-1 (with
spectral type A0V). For TYC 8378-252-2 (with spectral type F4/5V), the spectral range from 4360 to 4800
{\AA} containing Mg II (4384, 4481, and 4668) and Fe I (4383) lines was used for the RV calculations.
The spectra of HR 6031 and HR 3383 were used as a template for deriving the  RVs of the binary components.
The calculated RVs obtained from the cross-correlation method are depicted in Figs. \ref{FigRV_TYC5378}
and \ref{FigRV_TYC8378}  and are listed in Tables~\ref{tableA1} and \ref{tableA2} in Appendix \ref{secA}
for TYC 5378-1590-1 and TYC 8378-252-1, respectively.

\subsection{Photometry}
\label{photometry}

Our spectroscopic data were obtained from our own long-term observing campaigns; instead, we drew the
photometric data primarily from open surveys, satellite missions, and other long-term databases. We
relied primarily on TESS satellite data \citep{2015JATIS...1a4003R} for TYC 5378-1590-1. We used the
{\tt{lightkurve}} package \citep{2018ascl.soft12013L}  to extract the photometry from the TESS
database. For TYC 8378-252-1, we based our light curve solution exclusively on ground-based ASAS-SN data
\citep{2014ApJ...788...48S, 2017PASP..129j4502K}.

The primary databases mentioned above were supplemented with older data from the CRTS
\citep{2017MNRAS.469.3688D}, NSVS \citep{2004AJ....127.2436W}, and ASAS \citep{2002AcA....52..397P}, as
well as the discovery data for TYC 5378-1590-1 \citep{2013IBVS.6100...12M}, as inputs for our analysis.

\section{Analysis}
\label{analysis}

In view of the large TESS pixels (especially problematic in dense stellar fields with many neighboring
stars appearing on a pixel together with the target star), coupled with artificial instrumental trends
in the TESS photometry, a flattening method has to be applied to TESS data, and great caution needs to
be exercised when interpreting these data and conducting consequent  modeling. However, the TESS light
curves constitute the best available photometric data for this system by a long margin.

The {\sc PHOEBE package}, version 0.32svn \citep{2005ApJ...628..426P}, which was originally based on the
Wilson-Devinney code \citep{1971ApJ...166..605W} and applies Roche geometry, was used for our analysis
of the photometric data. By virtue of the good phase coverage of both the photometric and radial
velocity data for both our systems, our modeling procedure generated realistic physical parameters for
both of the eclipsing components in each system, and for their mutual orbits. Only a few additional
assumptions were required. The principal assumptions related to the gravity-brightening coefficients
and the albedos for both stars, together with the assumption of a pseudo-synchronization of rotation in
periastron for both systems; the obtained spectra were too noisy to be used for deriving the actual
rotational velocities of the components.

 \begin{figure}
   \centering
   \includegraphics[width=0.49\textwidth]{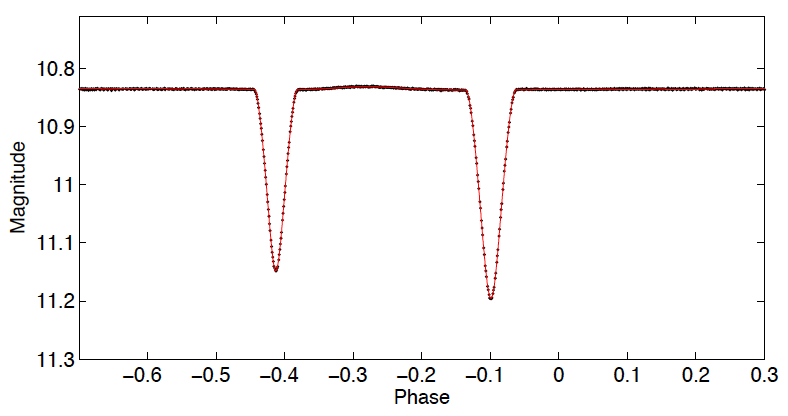}
   \caption{Light curve of TYC 5378-1590-1 based on TESS data and the solution provided with PHOEBE. The black crosses are TESS data and the red line is the PHOEBE solution.}
   \label{FigLC_TYC5378}
  \end{figure}

  \begin{figure}
   \centering
   \includegraphics[width=0.49\textwidth]{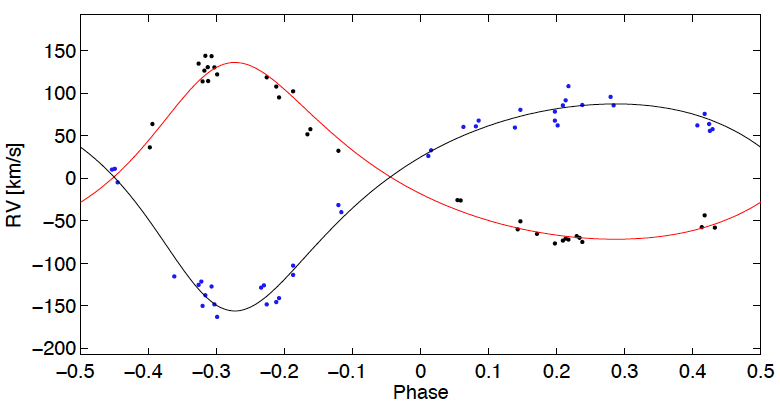}
   \caption{Radial velocity curves of TYC 5378-1590-1. The red curve applies to the primary component; the black curve applies to the secondary component.}
   \label{FigRV_TYC5378}
  \end{figure}

  \begin{figure}
   \centering
   \includegraphics[width=0.49\textwidth]{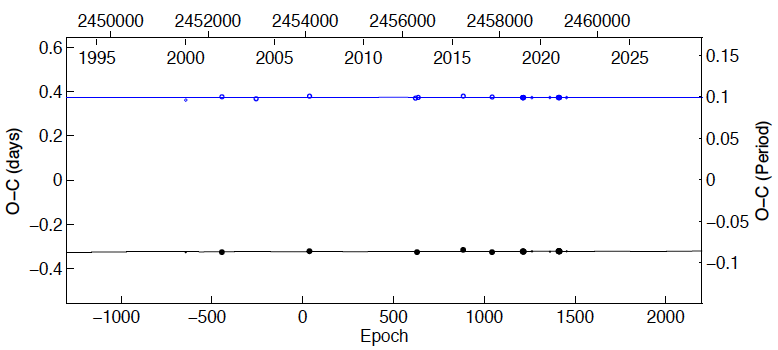}
   \caption{O-C diagram for the system TYC 5378-1590-1. The black symbols and     curve denote the primary component; the blue symbols and   curve denote the secondary component. A slow change in the omega angle is only barely noticeable.}
   \label{OCdiagrTYC5378}
  \end{figure}

    \begin{figure}
   \centering
   \includegraphics[width=0.49\textwidth]{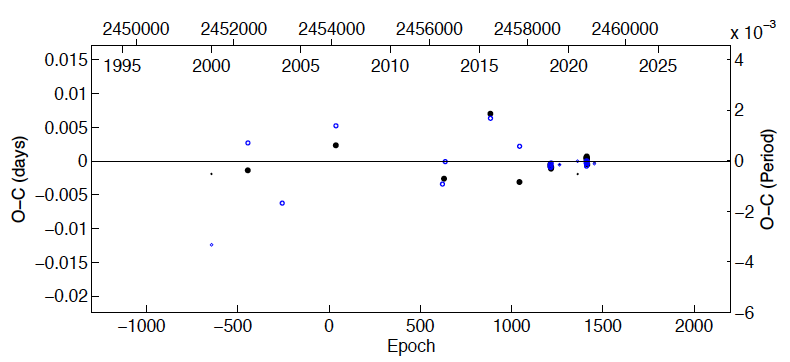}
   \caption{O-C diagram of residuals for the system TYC 5378-1590-1 after subtracting the apsidal motion fit.}
   \label{OCdiagrTYC5378resid}
  \end{figure}

\begin{table}
  \caption{Parameters obtained from the LC+RV fitting of TYC 5378-1590-1 and TYC 8378-252-1.}
  \label{LCRVparam}
  \centering
  \small
\begin{tabular}{lcc}
\hline \hline \multicolumn{1}{l}{Parameter} & \multicolumn{1}{c}{TYC 5378-1590-1} &
\multicolumn{1}{c}{TYC 8378-252-1} \\
\hline \hline
$T_0$ (HJD) & 2453946.517       & 2451981.779 \\
$P$ (d)     & 3.7323493         & 2.8776874 \\
\hline
$A$ (R$_{\odot}$) & 15.88 $\pm$ 0.07& 11.39 $\pm$ 0.15 \\
$V\gamma$ (km/s) & 1.38 $\pm$ 0.45  & 46.8 $\pm$ 1.4 \\
$e$         & 0.308 $\pm$ 0.037     & 0.193 $\pm$ 0.012 \\
$\omega$ (deg) & 18.0 $\pm$ 7.1     & 142.7 $\pm$ 8.3 \\
$q=M_2/M_1$ & 0.853 $\pm$ 0.008     & 1.14 $\pm$ 0.04 \\
$i$ (deg)   & 85.52 $\pm$ 0.06      & 87.08 $\pm$ 0.26 \\
$T_1$ (K)   & 9800 (fixed)          & 6308 $\pm$ 209 \\
$T_2$ (K)   & 8824 $\pm$ 87         & 6600 (fixed) \\
$\Omega_1$  & 9.344 $\pm$ 0.019     & 8.637 $\pm$ 0.036 \\
$\Omega_2$  & 9.549 $\pm$ 0.015     & 8.421 $\pm$ 0.062 \\
$r_1$ (mean)    & 0.16 $\pm$0.01    & 0.14 $\pm$0.01 \\
$r_2$ (mean)    & 0.22 $\pm$0.01    & 0.16 $\pm$0.01 \\
$L_1$ (TESS)    & 0.55 $\pm$0.02    & -- \\
$L_2$ (TESS)    & 0.33 $\pm$0.02    & -- \\
$L_3$ (TESS)    & 0.12 $\pm$0.05    & -- \\
$L_1$ (ASAS-SN g)   & --            & 0.43 $\pm$0.03 \\
$L_2$ (ASAS-SN g)   & --            & 0.57 $\pm$0.03  \\
$x_1$, $y_1$    & 0.78, 0.27        & 0.55, 0.22 \\
$x_2$, $y_2$    & 0.59, 0.29        & 1.18, 0.23 \\
\hline
\end{tabular}
\end{table}

\begin{table}
    \caption{Apsidal motion parameters of TYC 5378-1590-1 and TYC 8378-252-1.}
    \label{am_parameters}
  \centering
        \small
            \begin{tabular}{lcc}
            \hline \hline
            \multicolumn{1}{c}{Parameter} & \multicolumn{1}{c}{TYC 5378-1590-1} & \multicolumn{1}{c}{TYC 8378-252-1}  \\
            \hline \hline
$T_0$ (HJD)   & $2453946.517 \pm0.005$  & $2451981.779 \pm0.016$ \\
$P$ (d)       & $3.7323493 \pm0.0000468$    & $2.8776874 \pm0.0000101$ \\
$e$           & $0.308 \pm0.037$        & $0.193 \pm0.012$    \\
$\omega$ (deg)& $18 \pm7$           & $143 \pm8$ \\
$\dot{\omega}$  (deg yr$^{-1}$) & $0.036 \pm0.011$  & $0.362 \pm0.108$ \\
$U$ (yr)    & 10000 $\pm3056$           & $995 \pm297$ \\
            \hline
            \end{tabular}
\end{table}

\subsection{TYC 5378-1590-1}
\label{tyc5378}

TYC 5378-1590-1 (= 2MASS J06454395-0850355 = TIC 120023539, RA 06$^h$ 45$^m$ 43.95$^s$ DE -08$^\circ$
50$^\prime$ 35.50$^{\prime\prime}$, V = 10.9~mag) is a known eccentric eclipsing system. However, its
properties were  derived  very approximately in previous studies. The system was first identified as
an eccentric eclipsing system with a correct orbital period by \cite{2013IBVS.6100...12M}.
The system was included in our previous study \citep{2018A&A...619A..85Z}, where we used the 
  poor photometric data from the discovery paper and no spectroscopy. In that study a very slow
rate of apsidal motion and an eccentricity of 0.32 were derived for the system.

Our present study addresses this system with far superior TESS data and a longer time span of photometric observations, and with the recent spectroscopy obtained as part of our long-term campaigns from 2019 to  2021. The results of our fitting procedure are listed in Table \ref{LCRVparam} and plotted in Figs. \ref{FigLC_TYC5378} and \ref{FigRV_TYC5378}. The fit of the TESS data is superb. The RV data are much noisier (for the reasons mentioned earlier), but still present valuable information for inclusion in our analysis.

The solution presented in Table \ref{LCRVparam} was derived from our combined fit using both LC and RV
data together and constitutes a statistically meaningful result. A significant contribution of third
light emerged from the TESS data, in addition to the fluxes from the two main components. This effect
was caused by the relatively large TESS pixels, which allowed nearby stars unrelated to the TYC
5378-1590-1 system to contribute flux to the same pixels as the main target. We cannot speculate
about any third body in the system since no variation in the $O-C$ diagram is visible (see below).

Three different methods were used to determine the effective temperatures of the components of TYC
5378-1590-1, a system that has not been widely covered in  the literature (e.g., it had no assigned spectral
type prior to this study). First,  the color index of $B-V = 0.01 \pm0.09$, which was derived from the 
\citet{Hog_etal_2000} study, shows that the spectral type of the system is close to A0. Second,  the
solution of the RV curves obtained in the present study corresponds to a mass of the primary component
of approximately 2 M$_{\odot}$, which is in agreement with a spectral type of A0. Third,  using the
line-matching method, especially for the H$\gamma$, H$\beta$, and Mg II 4481 lines in the observed
spectra of the components and the sample spectra of the early A-type main-sequence stars in the 
\citet{Gray_Corbally_2008} book, the spectral type of the system was estimated to be close to A0.
Based on these results, we adopted a temperature of $9800 \pm200$ K, corresponding to a spectral type
of A0V in the calibrated data of \citet{Pecaut_Mamajek_2013}. The effective temperature of the primary component, which has a higher mass according to the solution for the RV curves, was taken as 9800 K, and this value was fixed in the WD iterations, whereas the effective temperature of the secondary component was adjusted during iterations.

As we   discuss in Section \ref{evolution}, both eclipsing components appear to be normal stars
located on the main sequence, most likely of spectral types A0 and A3. The primary
component might be slightly smaller in volume than  expected for a typical A0V star, while the
secondary appears slightly hotter than a typical A3V star. The locations of the two components on the
main sequence correspond with the young ages implied by the presence of an eccentric orbit.

Using all available photometric data for the star TYC 5378-1590-1, we also performed a period analysis
using the longest possible time interval. The data added to those used in our former study
\citep{2018A&A...619A..85Z} were primarily composed of the TESS observations, supplemented by more
recent ASAS-SN data. However, our results indicate only a slight evolution of the apsidal angle
$\omega$, meaning that the apsidal motion occurs at a very slow rate (on the order of thousands of
years). No additional trend is visible in the residuals, which are perfectly symmetric in their
distribution (see Fig. \ref{OCdiagrTYC5378resid}).

\subsection{TYC 8378-252-1}
\label{tyc8378}

TYC 8378-252-1 (= ASAS J185951-4711.8, RA 18$^h$ 59$^m$ 51.29$^s$ DE -47$^\circ$ 11$^\prime$
47.59$^{\prime\prime}$, V = 11.2~mag) has been mentioned only briefly as an eclipsing binary system in
prior work based on survey data from CRTS and ASAS. It was also included in our previous study
\citep{2018A&A...619A..85Z}, where we derived an eccentricity of $e=0.167$ for its orbit, and slow
apsidal motion with $U = 558$~yr. A more detailed analysis of this star has not been published to date.

The star was checked in various databases for photometric indices or any other indication of its color
or effective temperature. Markedly different values were obtained from various sources, for example,
$T_{eff}=6640$~K \citep{2017AJ....154..259S}, $6790 \pm 318$~ K \citep{2019AJ....158...93B}, and
$7004~$K \citep{2006ApJ...638.1004A}. Similar discrepant values were found for its distance. For
example, \cite{2006ApJ...638.1004A} presented the star's distance as 51~pc, whereas GAIA EDR3
\citep{2020yCat.1350....0G} presents a distance of 670~pc.

Therefore, three methods (as for the temperature determination of TYC 5378-1590-1) were applied to
determine the surface temperature of TYC 8378-252-1. First, according to the color index of $B-V = 0.60
\pm0.18$ derived from the study of \citet{Hog_etal_2000}, both  components are thought to be late
F-type main-sequence stars. Second,  the solution of the RV curves obtained in this study corresponds to a
mass of the more massive (secondary) component of approximately 1.3 M$_{\odot}$, which implies a
spectral type close to F5V. Third, using the line matching method for the H$\gamma$, H$\beta$, Ca I
(4226), Mg II (4481), Fe I (4383, 5041), Fe II (4923), and Mg I (5167 - 5183) triplet lines,  and using the
(CH) G band (see Fig. \ref {sample_spectra} for TYC 8378-252-1) in the spectra observed at the
conjunction phases of the secondary component and the sample spectra of the middle F-type main-sequence
stars in the book by \citet{Gray_Corbally_2008}, the spectral type of the secondary component was estimated
as F5V. Based on these results, we assigned a temperature corresponding to a spectral type of F5V,
equal to $6600 \pm200$ K, according to the calibrated data of \citet{Pecaut_Mamajek_2013}. The
preliminary solution of the ASAS-SN light curves indicated that the light contribution of the secondary
component is dominant. The secondary component also has a higher mass according to the solution of the
RV curves. Therefore, the temperature of the secondary component ($T_2$) was fixed at 6600 K, and the
temperature of the primary component ($T_1$) was adjusted in the iterations.

\begin{figure}
   \centering
   \includegraphics[width=0.49\textwidth]{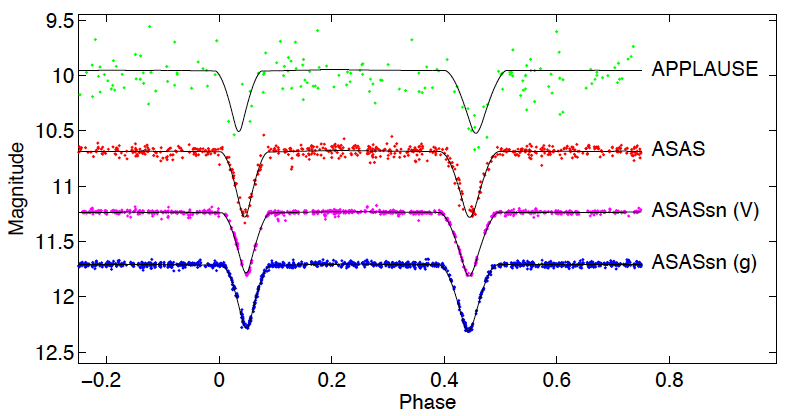}
   \caption{Light curves of TYC 8378-252-1.}
   \label{FigLC_TYC8378}
  \end{figure}

\begin{figure}
   \centering
   \includegraphics[width=0.49\textwidth]{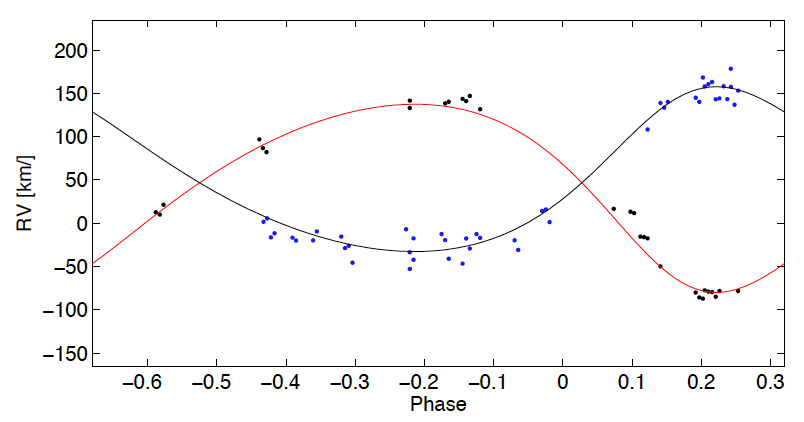}
   \caption{Radial velocity curves of TYC 8378-252-1. The red curve applies to the primary component; the black curve applies to the secondary component.}
   \label{FigRV_TYC8378}
  \end{figure}

No TESS photometry for this system is yet available (TESS will observe the system in July 2023), and
the existing photometry is of low quality. However, we used the best available photometric data
obtained by the ASAS-SN survey in its $g$ filter for a photometric analysis. We fitted these data using
PHOEBE together with the radial velocities (29 primary and 46 secondary velocities), resulting in the
solution presented in Table \ref{LCRVparam}. The fits are plotted in Figs. \ref{FigLC_TYC8378} and
\ref{FigRV_TYC8378}.

\begin{figure}
  \centering
  \includegraphics[width=0.49\textwidth]{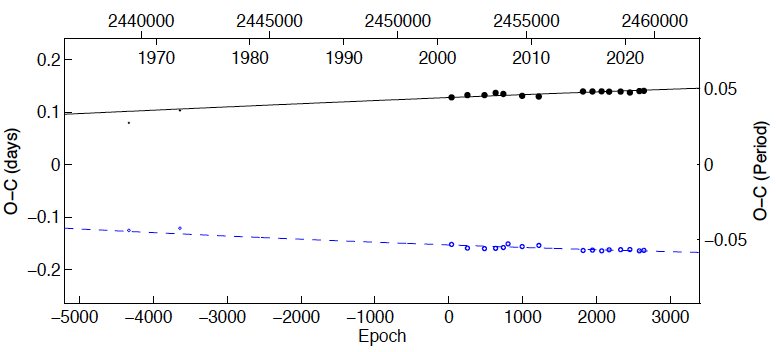}
  \caption{O-C diagram for the system TYC 8378-252-1.}
  \label{OCdiagrTYC8378}
\end{figure}

Our analysis of long-term period changes was conducted using all available photometric data.
Fortunately, the star was also found in the database of old photographic plates from German
observatories named APPLAUSE, available online,\footnote{https://www.plate-archive.org/query/} where
the star was observed during the 1960s and 1970s. From that photometry we were able to derive (albeit
with poor precision) the times of several eclipses. These data are highly important for constraining the
value of the long apsidal period with a higher degree of statistical confidence. In total, we collected
19 primary and 19 secondary times of eclipses for the analysis, covering a time span of more than 50
years. However, as seen from the $O-C$ diagram in Fig. \ref{OCdiagrTYC8378}, the change in the omega
angle is still very small and, accordingly, the apsidal motion is very slow. Our estimation of the
period of apsidal motion is on the order of 1000 yr. Our resulting eccentricity value is 0.193, only
slightly different from the value that was published by \cite{2014MNRAS.441..343S}, who presented the
system's orbital eccentricity as $0.17 \pm 0.065$, and from our former value published in
\cite{2018A&A...619A..85Z}.

\begin{figure}
  \centering
  \includegraphics[width=0.49\textwidth]{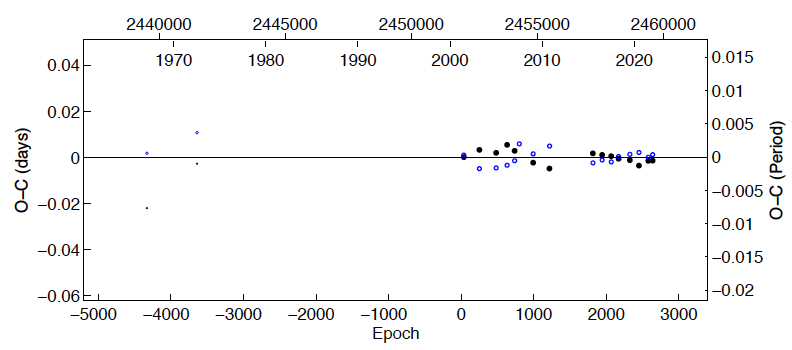}
  \caption{O-C diagram of residuals for the system TYC 8378-252-1 after subtracting the apsidal motion fit.}
  \label{OCdiagrTYC8378resid}
\end{figure}

\section{Discussion and conclusions}

In the present era of large-scale photometric and spectroscopic surveys, the question arises of whether a
dedicated study of one particular system is  sufficient  to produce results of
astrophysical importance. We answer in the affirmative. Currently, of the more than two million   known
eclipsing binary systems \citep{2006SASS...25...47W, 2022arXiv221100929M}, only 288 systems have had
their physical parameters derived to better than 2\% accuracy \citep{2015ASPC..496..164S}. Only a small
proportion of these systems are eccentric systems. Additional statistics to augment the small
present sample of eccentric systems, in the form of the P-e diagram that depicts the progress of
circularization in binary systems, would provide much-needed insight into the physical processes
occurring in stellar interiors, and into the orbital dynamics. We are therefore confident that this
study makes an important contribution to building the statistics of well-studied eccentric systems
showing apsidal motion.

\subsection{Absolute parameters}
\label{absolute_parameters}

The absolute parameters of binary stars are generally obtained by combining the information contained
in spectroscopic and photometric data and using the appropriate laws of physics. As a first step, the
masses of the components are calculated from Kepler's third law. The fractional radii of the components
($r_1$, $r_2$), obtained from the photometric data, are used to derive the absolute radii ($R_1$,
$R_2$). Subsequently, surface gravities ($g_1$, $g_2$) are derived directly. Using the effective
temperatures listed in Table \ref{abs_par}, we computed the bolometric magnitudes ($M_{\rm bol}$) and
luminosities ($L$) of the component stars of the two systems considered in the present study. We used
the nominal set of solar values   adopted by IAU 2015 Resolutions B2 and B3 as input in our
calculations. The absolute visual magnitudes $M_V$ were derived from the bolometric correction formula,
$BC$ = $M_{\rm bol}$ -- $M_{V}$. Bolometric corrections for the components were taken from the online
version of the color tabulation of \citet{Pecaut_Mamajek_2013}, according to their effective
temperatures.

The distance to each system was calculated from the well-known adaptation of Pogson's law: $M_{V}$ =
$m_{V}$ + 5 -- 5log$(d)$ -- $A_{V}$. The interstellar absorption and intrinsic color index were
computed using the following method. First, the total absorption towards the system in the galactic
disk in the $V$ band, $A_{\infty}(V)$, was taken from \citet{Schlafly_Finkbeiner_2011}, using the NASA
Extragalactic Database\footnote{ http://ned.ipac.caltech.edu/forms/calculator.html}. Second, the
interstellar absorption corresponding to the distance to the system, $A_{d}(V)$, was derived from the
formula given by \citet{Bahcall_Soneria_1980} (in their Eq.~8), using the system's Gaia-DR3 parallax
\citep{Gaia_2022}. Finally, the color excess for the system at the distance $d$ was estimated as
$E_{d}($B$-$V$)$ = $A_{d}(V)$/3.1.

The distance to TYC 5378-1590-1 (with corrections for interstellar absorption) was computed as 728
$\pm$60 pc using the distance modulus. When interstellar absorption was ignored, the distance was
computed as 1110 $\pm$80 pc. The photometric parallax formulation of \citet{Budding_Demircan_2007}
produces a distance of $1050 \pm80$ pc. However, considering the distance provided by Gaia-DR3
\citep{Gaia_2022}, 1292 $\pm$39 pc, the interstellar absorption for TYC 5378-1590-1 thus appears to be
overestimated. This would be in agreement with only low extinction towards the star as derived by
\cite{2020AJ....159...84B}. The  different effective temperatures of the component stars    play
a marginal role.

For TYC 8378-252-1, the distance was calculated as $550 \pm$40 pc when interstellar absorption was
taken into account, and  was calculated as 580 $\pm$50 pc when interstellar absorption was
ignored. The photometric parallax formulation of \citet{Budding_Demircan_2007} produces a distance of
$490 \pm60$ pc. The Gaia-DR3 \citep{Gaia_2022} distance for this star is provided as 670 $\pm$10 pc,
which is somewhat larger than the distance values calculated from the photometric data. As for the
previous case, we can speculate about slightly incorrect extinction or, on the contrary, our results  
may be affected by larger errors than presented.

The absolute parameters of TYC 5378-1590-1 and TYC 8378-252-1 as determined by our analysis, together
with their errors, are listed in Table \ref{abs_par}.

\begin{table*}
  \centering
\caption{Absolute parameters of TYC 5378-1590-1 and TYC 8378-252-1.} \label{abs_par}
\begin{tabular}{lllll}
\hline \hline
& \multicolumn{2}{c}{TYC 5378-1590-1} & \multicolumn{2}{c}{TYC 8378-252-1}  \\
Parameter       & Primary       & Secondary & Primary       & Secondary     \\
\hline \hline
$a$ (R$_{\odot}$)       & \multicolumn{2}{c}{15.88 $\pm$0.07}   & \multicolumn{2}{c}{11.39 $\pm$0.15} \\
$M$ (M$_{\odot}$)       & $2.09 \pm0.06$    & $1.78 \pm0.05$    & $1.12 \pm0.06$    & $1.28 \pm0.11$ \\
$R$ (R$_{\odot}$)       & $1.96 \pm0.05$    & $1.69 \pm0.06$    & $1.58 \pm0.07$    & $1.80 \pm0.06$ \\
log $\{g$ (cm s$^{-2}$)$\}$     & $4.17 \pm0.06$    & $4.23 \pm0.07$    & $4.09 \pm0.07$    & $4.04 \pm0.08$ \\
$T$ (K)             & $9800 \pm200$ & $8824 \pm180$ & $6308 \pm210$     & $6600 \pm200$ \\
log $\{L$ (L$_{\odot}$)$\}$     & $1.50 \pm0.06$    & $1.19 \pm0.07$    & $0.55 \pm0.10$    & $0.74 \pm0.08$ \\
$M_{\rm bol}$ (mag)         & $0.98 \pm0.14$    & $1.76 \pm0.17$    & $3.36 \pm0.24$    & $2.88 \pm0.20$ \\
$M_{V}$ (mag)           & $1.21 \pm0.14$    & $1.83 \pm0.17$    & $3.39 \pm0.24$    & $2.90 \pm0.20$ \\
$E$($B$-$V$) (mag)  & \multicolumn{2}{c}{0.30}      & \multicolumn{2}{c}{0.04} \\
$B$-$V$ (mag)       & \multicolumn{2}{c}{0.01 $\pm$0.09$^{a}$}  & \multicolumn{2}{c}{0.60 $\pm$0.18$^{a}$} \\
$V$ (mag)           & \multicolumn{2}{c}{10.95 ~$\pm$0.07$^{a}$}    & \multicolumn{2}{c}{11.18 ~$\pm$0.12$^{a}$} \\
$M_{V}$ $(system)$ (mag)    & \multicolumn{2}{c}{$0.72 ~\pm$0.17}   & \multicolumn{2}{c}{$2.37 ~\pm$0.18} \\
$d$ (pc)                & \multicolumn{2}{c}{728 $\pm$60}       & \multicolumn{2}{c}{550 $\pm$40} \\
$d_{Gaia-DR3}$(pc)  & \multicolumn{2}{c}{1292 $\pm$39$^{b}$}    & \multicolumn{2}{c}{670 $\pm$10$^{b}$} \\
\hline
\end{tabular}
\begin{flushleft}
\item \textit{Notes:} $^{a}$ \citet{Hog_etal_2000};  $^{b}$ \citet{Gaia_2022}.
\end{flushleft}
\end{table*}

\subsection{Evolutionary status}
\label{evolution}

The Granada \citep[e.g.,][]{Claret_2006} and Padova \citep[e.g.,][]{Bressan_etal_2012} evolution models
were used to study the evolutionary status of these two eccentric binary systems, using the physical
parameters listed in Table \ref{abs_par}. The log $T$ versus log $g$ diagram and the log $T$ versus log $L$
diagram (i.e., the H-R diagram) were used to determine the metallicity values $Z$ of the component stars,
whereas the log (age) versus radius diagram and the H-R diagram were used to determine their ages.

\begin{figure}
\centering
    \includegraphics[scale=1.2]{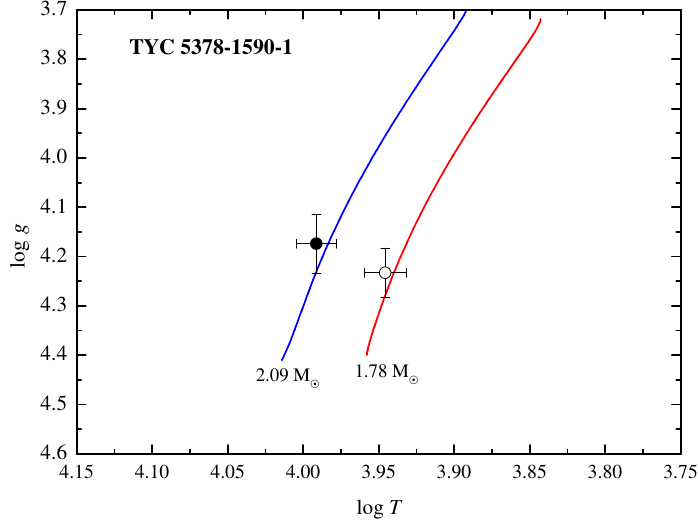} \\
    \includegraphics[scale=1.2]{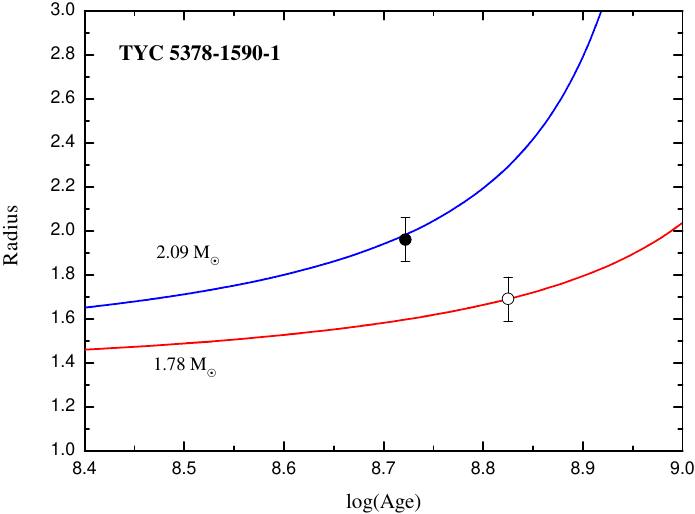} \\
    \includegraphics[scale=1.2]{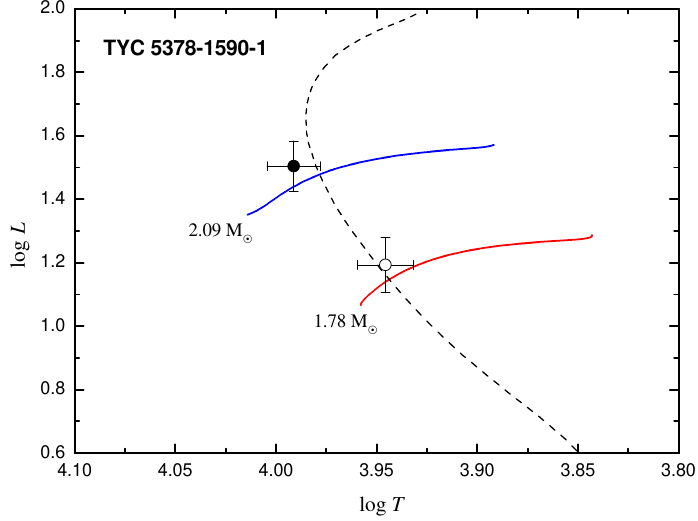}
\caption{ Location of the components of TYC 5378-1590-1 in the log ($T_{\rm eff}$)--log ($g$) diagram
(upper panel), log (age)--radius diagram (middle panel), and H-R diagram (lower panel). The Granada
evolutionary tracks from ZAMS to TAMS \citep{Claret_2006} for stars of 2.09 M$_{\odot}$ (blue
line) and 1.78 M$_{\odot}$ (red line), corresponding to the primary and secondary stars,
respectively, are displayed for $Z=0.010$ in all panels. The middle panel displays the radial evolution
of the component stars. As discussed in the text, the age of TYC 5378-1590-1 was estimated as
$537\pm55$ Myr. The Padova isochrone line for an age of 537 Myr \citep{Bressan_etal_2012} is indicated
by the dashed black curve superimposed on the H-R diagram. In all diagrams, filled and open circles
 represent the primary and secondary components, respectively. The vertical and horizontal lines are the
error bars of the measured quantities. } \label{evol1}
\end{figure}

\begin{figure}
\centering
\includegraphics[scale=1.2]{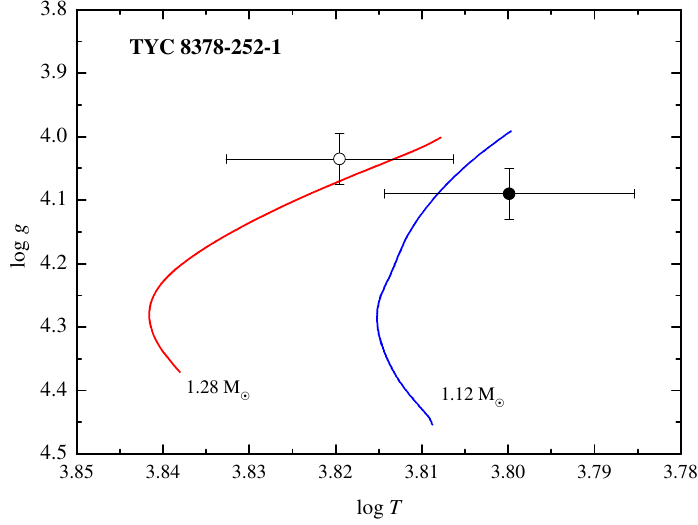}
\includegraphics[scale=1.2]{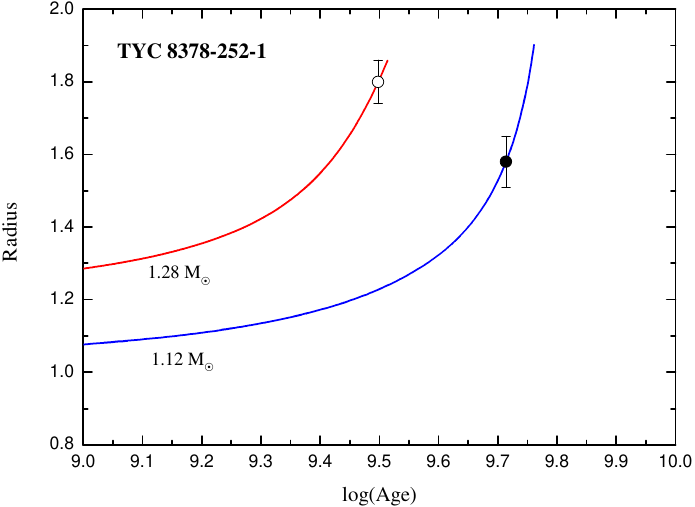}
\includegraphics[scale=1.2]{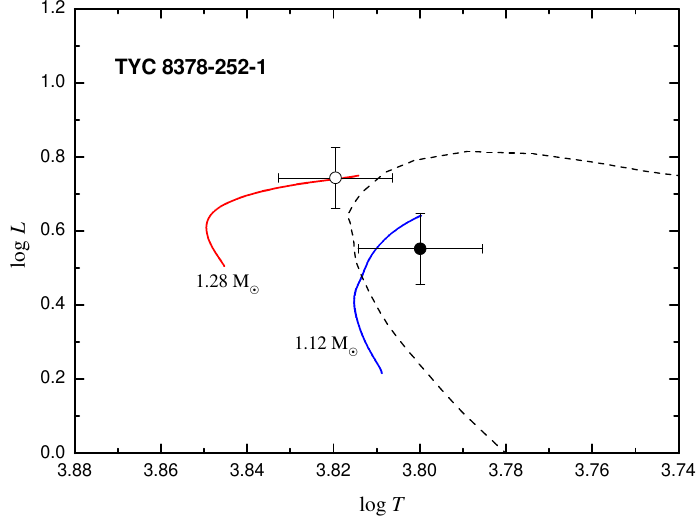}
\caption{ Location of the components of TYC 8378-252-1 in the log ($T_{\rm eff}$)--log ($g$) diagram
(upper panel), log (age)--radius diagram (middle panel), and H-R diagram (lower panel). The Granada
evolutionary tracks from ZAMS to TAMS \citep{Claret_2006} for stars of 1.12 M$_{\odot}$ (blue
line) and 1.28 M$_{\odot}$ (red line), corresponding to the primary and secondary stars,
respectively, are displayed for $Z=0.007$ in all panels. The middle panel displays the radial evolution
of the component stars. As discussed in the text, the age of TYC 8378-252-1 was estimated as $3.98
\pm0.80$ Gyr. The Padova isochrone line for an age of 3.98 Gyr \citep{Bressan_etal_2012} is indicated
by the dashed black curve superimposed on the H-R diagram. In all diagrams, filled and open circles
  represent the primary and secondary components, respectively. The vertical and horizontal lines are the
error bars of the measured quantities. } \label{evol2}
\end{figure}

In Figs. \ref{evol1} and \ref{evol2}, we display the evolutionary tracks of the component stars that
correspond to their measured masses. The positions of the component stars indicated in the log $T$ versus
log $g$ diagram and in the H-R diagram correspond to values of the metallicity $Z$ for TYC 5378-1590-1
and TYC 8378-252-1 of $Z=0.010\pm 0.001$ and $0.007\pm 0.002$, respectively.

The evolutionary progression of the radii of the components of TYC 5378-1590-1 in accordance with
their determined masses and $Z=0.010$ are displayed in the middle panel in Fig. \ref{evol1}. The
positions of the components are plotted according to their radii,  as listed in Table \ref{abs_par}. The
calculated radius of the primary component corresponds to a log(age)=8.72 and the calculated radius of
the secondary component corresponds to a log(age)=8.83. In the H-R diagram (lower panel
of Fig. \ref{evol1}), the positions of the components determined according to their calculated physical
parameters fit the isochrone of log(age)=8.73  within the error limits. We assigned the value
log(age)=$8.73\pm 0.07$ to TYC 5378-1590-1 (i.e., an age of $537\pm55$ Myr). This value for the age
implies that the theoretical model predicts a smaller radius than the observational radius of the
secondary component, as shown in the radius evolution diagram of the component stars (displayed in the
middle panel of Fig. \ref{evol1}).

The evolutionary progression of the radii of the components of TYC 8378-252-1 in accordance with
their determined masses and $Z=0.007$ are displayed in the middle panel in Fig. \ref{evol2}. The
positions of the components are plotted according to their radii, as listed in Table \ref{abs_par}. The
calculated radius of the primary component corresponds to a log(age)=9.71 and the calculated radius of
the secondary component corresponds to a log(age)=9.50. According to the H-R diagram in the lower panel
of Fig. \ref{evol2}, the positions of the components determined according to their calculated physical
parameters fit the isochrone of log(age)=9.60 within the error limits. We assigned the value
log(age)=$9.60\pm 0.15$ to TYC 8378-252-1 (i.e.,  an age of $3.98 \pm0.80$ Gyr). This value for the
age implies that the theoretical model predicts a smaller radius than the observational radius for the
primary component (the less massive component, as explained above), as shown in the
radius evolution diagram of the component stars (displayed in the middle panel of Fig. \ref{evol2}).

\subsection{Internal structure constant}
\label{isc}

In this section we describe the results of our calculations of the internal structure constants of
these two eccentric binary systems, using the method and equations given by \citet{Claret_etal_2021}
and references therein.

The observational apsidal motion rates $\dot{\omega}_{obs}$ were derived from the $O-C$ analysis in
Section \ref{analysis}, as $0.000368 \pm0.000112$  deg cycle$^{-1}$ for TYC 5378-1590-1 and $0.002852
\pm0.000851$  deg cycle$^{-1}$ for TYC 8378-252-1.

The relativistic term in the rate of apsidal motion was calculated using the following equation, which
was described by \citet{Levi-Civita_1937} and presented by \citet{Gimenez_1985}:
\begin{equation}
\dot{\omega}_{rel}=5.447\times10^{-4}\dfrac{(M_{1}+M_{2})^{2/3}}{(1-e^{2})P^{2/3}} \qquad
(\textrm{deg\:cycle$^{-1}$}). \label{omega_rel}
\end{equation}
Here $M_{1}$ and $M_{2}$ represent the masses of the component stars in solar units, $e$  
the orbital eccentricity, and $P$   the orbital period in days. Subsequently, the relativistic
terms were calculated as $0.000615 \pm0.000056$ deg cycle$^{-1}$ for TYC 5378-1590-1 and $0.000500
\pm0.000044$ deg cycle$^{-1}$ for TYC 8378-252-1.

There is a remarkable result here for TYC 5378-1590-1. If we extract the classical term
$\dot{\omega}_{cl}$ from the well-known equation
$\dot{\omega}_{obs}$=$\dot{\omega}_{cl}$+$\dot{\omega}_{rel}$, the term $\dot{\omega}_{cl}$   takes a
negative value since $\dot{\omega}_{rel} > \dot{\omega}_{obs}$. This  scenario is often encountered in
eclipsing binary systems with a very slow rate of apsidal motion \citep[e.g.,][]{Gimenez_1985}. The most
famous of these examples is the system DI Her  \citep[e.g.,][]{Guinan_Maloney_1985}.
\citet{Albrecht_etal_2009} observationally demonstrated that this phenomenon in DI Her is caused by
misaligned spin and orbital axes of the component stars (see also \citealt{1985SvAL...11..224S}). We
were not able to perform a similar procedure in order to show whether there is spin-axis misalignment
in the components of TYC 5378-1590-1 because of our lack of access to high-resolution spectra taken at
the conjunction phases of this system.

The classical term was derived as $\dot{\omega}_{cl}=0.002352 \pm0.000732$ deg cycle$^{-1}$ for TYC
8378-252-1, contributing approximately 82$\%$ of the observed rate of apsidal motion. Using the absolute
parameters   listed in Table \ref{abs_par}, the mean observational internal structure constant of this
system was calculated as $\log{\bar{k}_{2,obs}}=-2.728 \pm{0.120}$.

The theoretical internal structure constants $\bar{k}_{2,theo}$ were predicted from the Granada
evolution models \citep{Claret_Gimenez_1992,Claret_2006}. The Granada models produce theoretical values
of $\log{k_{2}}$ for a grid of masses, surface gravities, luminosities, effective temperatures, ages,
and metallicities. Values of $k_{2i}$ for each component star were calculated by interpolating between
the model values and the observed physical properties listed in Table \ref{abs_par}.

The predicted metallicities and ages of TYC 5378-1590-1 and TYC 8378-252-1 based on our analysis of
Granada models are summarized in Table \ref{ics_par}. These results were also checked against the
Padova and Geneva models. Using the absolute parameters given in Table \ref{abs_par} (in accordance with
the masses of the components and the values of $Z=0.010$ and 0.007), interpolating, taking the weighted
average and then the logarithm, the mean observational internal structure constants of TYC 5378-1590-1
and TYC 8378-252-1 were determined as $\log{\bar{k}_{2,theo}}=-2.435 $ and $-2.424 $, respectively,
from the Granada Models. These values are given in Table \ref{ics_par} for ease of comparison of the
observational and theoretical values.

\begin{table}
  \centering
\caption{Metallicities, observational and theoretical internal structure constants, and evolutionary
and circularization ages of TYC 5378-1590-1 and TYC 8378-252-1.} \label{ics_par}
\begin{tabular}{lll}
\hline 
Parameter & TYC 5378-1590-1 & TYC 8378-252-1  \\
\hline 

$Z$                     & 0.010 $\pm$0.001      & 0.007 $\pm$0.002 \\
log $\bar{k}_{2,obs}$       & ---                   & $-2.728 \pm0.120$  \\
log $\bar{k}_{2,theo}$      & $-2.435$          & $-2.424$ \\
Age (yr)                    & $(5.37 \pm0.55) \times 10^8$      & $(3.98 \pm0.80) \times 10^9$  \\
$\tau_{circ}$ (yr)          & $1.10 \times 10^{11}$ & $2.76 \times 10^8$ \\
\hline
\end{tabular}
\end{table}

For TYC 5378-1590-1, log $\bar{k}_{2,obs}$ could not be calculated because the term $\dot{\omega}_{cl}$
is negative, as mentioned before, and therefore this parameter could not be entered in Table \ref{ics_par}.
As an approach to solving this problem, we started by accepting the expected result (i.e., log
$\bar{k}_{2,obs}$ = log $\bar{k}_{2,theo}$) and used the equations given by \citet{Claret_etal_2021} to
calculate the classical apsidal motion rate as $\dot{\omega}_{cl}$ = 0.00199 deg cycle$^{-1}$ (hence,
the period of apsidal motion as $U = 1410$ yr). Upon entering these values as input values to the $O-C$
analysis and repeating the iterations, we obtained $\dot{\omega}_{obs} = 0.00261 \pm0.00091$ deg
cycle$^{-1}$ (and $U = 1409.5 \pm361.7$ yr). Calculations of other parameters resulted in the same
values as those listed in Table \ref{am_parameters}, within the error limits. Thus, in the $O-C$
analysis of TYC 5378-1590-1, two different models appeared for the apsidal motion of the system: Model
A with a long period of apsidal motion ($U=9937$ yr, $rms = 0.0023$), and Model B with a short period
of apsidal motion ($U=1410$ yr, $rms = 0.0064$). In the comparison of these two models, while the $rms$
values point to Model A as the most likely solution, the negative result of the classical term in this
model creates a problem similar to the situation in DI Her. However, the fact that the $rms$ values
between these two models are not very different in terms of statistical significance prevents a
definitive determination of the most realistic model from being made. In addition, the fact that the periods
of apsidal motion in both models are on the order of a thousand years implies that a sufficient number of
eclipse times could only be obtained in a few centuries to indicate which model is correct. Therefore,
the solution to this problem for TYC 5378-1590-1 remains open-ended for the time being.

For TYC 8378-252-1, the value of $\log{\bar{k}_{2,theo}}$ predicted from the Granada models  was
approximately 11$\%$ larger than the observed value (that is, by $1\sigma$). In comparing the values of
$\log{\bar{k}_{2,obs}}$ and $\log{\bar{k}_{2,theo}}$, the absolute parameters derived from the
observations are expected to be compatible with those of the evolution models and to indicate the same age
for both components of a given binary star. In the case of TYC 8378-252-1 (see Sect.
\ref{evolution}), in Fig. \ref{evol2}, for the less massive primary component, the upper and lower
panels show that the evolution model predicts a temperature value greater than the observational value
within the margins of error, while the middle and lower panels show that the evolution model
predicts a smaller radius than the observational value. Although the age determination in the H-R
diagram is accepted as log(age)=9.60 within the error limits, the middle panel also shows that the
measured radii of the component stars indicate different ages for the system.

\citet{Lester_Gies_2018} pointed out that the mismatch between the observational and theoretical radii
of these component stars (which translates to incompatible ages for the components) is encountered for
for evolved F-type eclipsing binary components similar to TYC 8378-252-1. As \citet{Lester_Gies_2018} mentioned, convective cores have started developing in evolved F-type stars, given their mass in the range of 1.1--1.7 Msun, and given that convective core overshooting disturbs central condensation. This situation is not adequately represented in current evolution models. For example, in the case of the TYC 8378-252-1
system, the evolution model gives an internal structure constant larger than the observational value,
which indicates that the components are more centrally condensed than predicted by models.

The evolutionary age of TYC 8378-252-1, which has a remarkably eccentric orbit ($e \sim 0.2$), is quite
large, and both components appear to have evolved almost to the TAMS as main-sequence stars. This casts
doubt on whether the expected orbit-circularization mechanism has been operating in this system. To
investigate this problem, the circularization timescale was calculated using the \citet{Zahn_1977}
theory.

Considering the spectral types (and effective temperatures) of the components of TYC 8378-252-1 (see
Section \ref{tyc8378}), it can be accepted that these stars have convective envelopes. In this case,
using Eq. (6.2) in \citet{Zahn_1977} for stars with convective envelopes and the corresponding absolute
parameters in Table \ref{abs_par} (above), the circularization timescale was calculated as $\tau_{circ}
= 2.76 \times 10^{8}$ yr. Comparing the results in Table \ref{ics_par}, since the evolutionary age of
TYC 8378-252-1 is approximately ten times older than its circularization age, this binary system would
  be expected to have a circular orbit at present. Therefore, our results are not compatible with
the predictions of the \citet{Zahn_1977} theory of circularization.

The spectral types of the components of TYC 5378-1590-1 (see Section \ref{tyc5378}) indicate that these
stars should have convective cores and radiative envelopes. Therefore,  the \citet{Claret_Cunha_1997}
formulation of the \citet{Zahn_1977} theory for stars with convective cores and radiative envelopes
was used to find the circularization timescale of TYC 5378-1590-1. The tidal torque constant $E_2$ and
fractional gyration radius $R_{gyr}$  in the formulas were determined by interpolating, according to
the mass and surface gravity of the component stars, from the Granada evolution models
\citep{Claret_2006}. Using $\log (E_{21}) = -7.008$, $R_{gyr,1}=0.198$ and $\log (E_{22}) = -7.557$,
$R_{gyr,2}=0.199$ for the primary and secondary components of TYC 5378-1590-1 and the absolute
parameters listed in Table \ref{abs_par}, the equations of \citet{Claret_Cunha_1997} predict the
circularization timescale to be $\tau_{circ} = 1.10 \times 10^{11}$ yr. As a result, the
circularization age of TYC 5378-1590-1 is approximately 200 times older than its evolutionary age and
the orbit of the system is still eccentric, which shows that the observational findings for this system
agree with the  \citet{Zahn_1977} theory of circularization.

\begin{acknowledgements}

This research was supported by T\"{U}B\.{I}TAK (Scientific and Technological Research Council of
T\"{u}rkiye) under Grant No. 121F203. We would like to thank the Time Allocation Committee of the SAAO
for ample observing time. CE and FM thank the National Research Foundation of South Africa and the
University of Johannesburg for funding. Funding for APPLAUSE has been provided by DFG (German Research
Foundation, Grant), Leibniz Institute for Astrophysics Potsdam (AIP), Dr. Remeis Sternwarte Bamberg
(University Nuernberg/Erlangen), the Hamburger Sternwarte (University of Hamburg) and Tartu
Observatory. Plate material also has been made available from Th\"uringer Landessternwarte Tautenburg.

It is a pleasure to express our appreciation of the high quality and ready availability, via the
Mikulski Archive for Space Telescopes (MAST), of data collected by the TESS mission. Funding for the
TESS mission is provided by the NASA Explorer Program. The authors thank the ASAS, ASAS-SN, CRTS, NSVS
and TESS teams for making all of the observations easily public available. This research has made
partial use of data from the European Space Agency (ESA) mission {\it Gaia}
(\url{https://www.cosmos.esa.int/gaia}), processed by the {\it Gaia} Data Processing and Analysis
Consortium (DPAC, \url{https://www.cosmos.esa.int/web/gaia/dpac/consortium}). Funding for the DPAC has
been provided by national institutions, in particular the institutions participating in the {\it Gaia}
Multilateral Agreement. This research has partly made use of the SIMBAD and VIZIER databases, operated
at CDS, Strasbourg, France, and of NASA's Astrophysics Data System Bibliographic Services.

 \end{acknowledgements}




\begin{appendix}
 \section{Tables of radial velocity measurements of TYC~5378-1590-1 and TYC~8378-252-1} \label{secA}
 \FloatBarrier
\begin{table}[h!]
\caption{RV measurements, with standard errors, of the components of TYC~5378-1590-1.} \label{tableA1}
 \scriptsize
\begin{tabular}{cccccc}
        &           &            &              &             &            \\
\hline
No & Time      & $RV_{1}$ & $\sigma_{1}$ & $RV_{2}$ & $\sigma_{2}$ \\
   & (HJD-2450000) & (km s$^{-1}$)   & (km s$^{-1}$)        & (km s$^{-1}$)   & (km s$^{-1}$)    \\
\hline
1   &   8817.4829   &$  -25.87  $   &$  3.7 $   &$  -   $   &$  -   $\\
2   &   8817.5002   &$  -26.32  $   &$  4.72    $   &$  -   $   &$  -   $\\
3   &   8817.5154   &$  -   $   &$  -   $   &$  60.21   $   &$  13.2    $\\
4   &   8819.5261   &$  36.13   $   &$11.22 $   &$  -   $   &$  -   $\\
5   &   8819.5413   &$  63.56   $   &$13.21 $   &$  -   $   &$  -   $\\
6   &   8820.5616   &$  32.11   $   &$15.33 $   &$  -31.75  $   &$  6.68    $\\
7   &   8820.5772   &$  -   $   &$  -   $   &$  -40.1   $   &$  6.66    $\\
8   &   8822.5559   &$  -57.71  $   &$  3.94    $   &$  -   $   &$  -   $\\
9   &   8822.5714   &$  -43.87  $   &$  2.56    $   &$  75.6    $   &$  10.95   $\\
10  &   8823.5480   &$  113.78  $   &$19.71 $   &$  -150.38 $   &$  17.19   $\\
11  &   8823.5638   &$  143.84  $   &$16.25 $   &$  -137.82 $   &$  17.16   $\\
12  &   8823.5790   &$  114.15  $   &$  9.64    $   &$  -   $   &$  -   $\\
13  &   8825.5261   &$  -73.52  $   &$  9.16    $   &$  85.59   $   &$  11.22   $\\
14  &   8825.5412   &$  -71.14  $   &$  6.32    $   &$  91.51   $   &$  15.69   $\\
15  &   8825.5564   &$  -72.23  $   &$  5.55    $   &$  108.09  $   &$  15.36   $\\
16  &   9127.5835   &$  -   $   &$  -   $   &$  59.44   $   &$  8.26    $\\
17  &   9127.5986   &$  -60.21  $   &$  13.23   $   &$  -   $   &$  -   $\\
18  &   9127.6137   &$  -50.84  $   &$  12.3    $   &$  80.34   $   &$  10.87   $\\
19  &   9128.5847   &$  -   $   &$  -   $   &$  61.97   $   &$  29.81   $\\
20  &   9129.5795   &$  134.65  $   &$12.53 $   &$  -125.54 $   &$  15.42   $\\
21  &   9129.5947   &$  -   $   &$  -   $   &$  -121.62 $   &$  16.33   $\\
22  &   9129.6112   &$  126.44  $   &$18.25 $   &$  -   $   &$  -   $\\
23  &   9129.6296   &$  130.46  $   &$11.34 $   &$  -   $   &$  -   $\\
24  &   9152.5714   &$  51.51   $   &$17.20 $   &$  -   $   &$  -   $\\
25  &   9152.5868   &$  57.53   $   &$16.33 $   &$  -   $   &$  -   $\\
26  &   9155.5726   &$  -   $   &$  -   $   &$  -115.74 $   &$  14.23   $\\
27  &   9157.5631   &$  -65.56  $   &$  11.2    $   &$  -   $   &$  -   $\\
28  &   9214.4971   &$  -   $   &$  -   $   &$  55.53   $   &$  11.23   $\\
29  &   9214.5123   &$  -   $   &$  -   $   &$  57.53   $   &$  13.33   $\\
30  &   9217.3792   &$  -76.94  $   &$  15.4    $   &$  78.27   $   &$  3.55    $\\
31  &   9217.3943   &$  -   $   &$  -   $   &$  61.99   $   &$  9.18    $\\
32  &   9219.5826   &$  107.64  $   &$  5.73    $   &$  -145.91 $   &$  38.38   $\\
33  &   9219.5975   &$  94.93   $   &$  6.35    $   &$  -141.16 $   &$  26.63   $\\
34  &   9300.3373   &$  -   $   &$  -   $   &$  63.56   $   &$  13.2    $\\
35  &   9300.3685   &$  -58.23  $   &$17.75 $   &$  -   $   &$  -   $\\
36  &   9301.3390   &$  143.5   $   &$20.56 $   &$  -127.54 $   &$  24.3    $\\
37  &   9301.3542   &$  130.29  $   &$19.60 $   &$  -148.58 $   &$  22.3    $\\
38  &   9301.3695   &$  121.92  $   &$19.30 $   &$  -163.24 $   &$  30.89   $\\
39  &   9303.3432   &$  -68.24  $   &$  11.2    $   &$  -   $   &$  -   $\\
40  &   9303.3583   &$  -70.25  $   &$  10.9    $   &$  -   $   &$  -   $\\
41  &   9303.3738   &$  -75.07  $   &$  13.69   $   &$  86  $   &$  21.22   $\\
42  &   9305.3434   &$  -   $   &$  -   $   &$  -128.89 $   &$  22.96   $\\
43  &   9305.3587   &$  -   $   &$  -   $   &$  -126.24 $   &$  24.37   $\\
44  &   9305.3740   &$  118.58  $   &$17.37 $   &$  -148.57 $   &$  30.8    $\\
45  &   9306.2612   &$  -   $   &$  -   $   &$  26.09   $   &$  11.2    $\\
46  &   9306.2765   &$  -   $   &$  -   $   &$  32.78   $   &$  12  $\\
47  &   9307.2611   &$  -   $   &$  -   $   &$  95.53   $   &$  20.22   $\\
48  &   9307.2782   &$  -   $   &$  -   $   &$  85.56   $   &$  24.31   $\\
49  &   9308.2574   &$  -   $   &$  -   $   &$  10.1    $   &$  13.2    $\\
50  &   9308.2726   &$  -   $   &$  -   $   &$  10.96   $   &$  11.23   $\\
51  &   9308.2879   &$  -   $   &$  -   $   &$  -5.04   $   &$  8.29    $\\
52  &   9309.2512   &$  102.14  $   &$16.77 $   &$  -103.03 $   &$  14.76   $\\
53  &   9310.2542   &$  -   $   &$  -   $   &$  60.88   $   &$  13.33   $\\
54  &   9310.2696   &$  -   $   &$  -   $   &$  67.57   $   &$  13.5    $\\
\hline
\end{tabular} 
\end{table}

\begin{table}
\begin{center}
\caption{RV measurements, with standard errors, of the components of TYC~8378-252-1.}\label{tableA2}
 \scriptsize
\begin{tabular}{cccccc}
        &           &            &              &             &            \\
\hline
No & Time      & $RV_{1}$ & $\sigma_{1}$ & $RV_{2}$ & $\sigma_{2}$ \\
   & (HJD-2450000) & (km s$^{-1}$)   & (km s$^{-1}$)        & (km s$^{-1}$)   & (km s$^{-1}$)    \\
\hline
1   &   9060.3844   &$  -   $   &$  -   $   &   $-12.51$    &$  18.33   $\\
2   &   9060.3995   &$  138.49  $   &$17.00 $   &   $-19.48$    &$  13.44   $\\
3   &   9060.4146   &$  140.49  $   &$16.70 $   &   $-40.87$    &$  9.59    $\\
4   &   9060.5298   &$  -   $   &$  -   $   &   $-12.41$    &$  14.92   $\\
5   &   9060.5450   &$  131.80  $   &$  17.80$  &   $-16.73$    &$  14.76   $\\
6   &   9061.2945   &$  -49.78$ &$  18.31   $   &$  138.93  $   &$  12.93   $\\
7   &   9061.3100   &$  -   $   &$  -   $   &$  133.63  $   &$  8.35    $\\
8   &   9061.3251   &$  -   $   &$  -   $   &$  140.3   $   &$  11.37   $\\
9   &   9061.4786   &$  -77.54  $   &$  11.12$  &   $158.16$    &$  10.4    $\\
10  &   9061.4937   &$  -78.94  $   &$  10.13$  &   $160.84$    &$  11.2    $\\
11  &   9061.5091   &$  -79.61  $   &$  13.20$  &   $163.26$    &$  39.76   $\\
12  &   9061.5243   &$  -84.96  $   &$  14.40$  &   $143.31$    &$  14.05   $\\
13  &   9061.5397   &$  -78.27  $   &$  11.20$  &   $144.30$    &$  9.64    $\\
14  &   9062.5218   &$  -   $   &$  -   $   &   $   1.77$   &$  11.3    $\\
15  &   9062.5370   &$  -   $   &$  -   $   &   $   5.60$   &$  15.08   $\\
16  &   9062.5526   &$  -   $   &$  -   $   &   $-16.33$    &$  27.1    $\\
17  &   9062.5678   &$  -   $   &$  -   $   &   $-11.55$    &$  36.65   $\\
18  &   9063.3494   &$  143.84  $   &$15.23 $   &   $-46.67$    &$  11.96   $\\
19  &   9063.3646   &$  141.16  $   &$15.00 $   &   $-17.58$    &$  28.98   $\\
20  &   9063.3796   &$  147.18  $   &$16.23 $   &   $-29.13$    &$  38.68   $\\
21  &   9063.5659   &$  -   $   &$  -   $   &   $-19.62$    &$  24.23   $\\
22  &   9063.5810   &$  -   $   &$  -   $   &   $-30.87$    &$  21.31   $\\
23  &   9064.3187   &$  -80.00  $   &$  41.79$  &   $145.17$    &$  29.69   $\\
24  &   9064.3338   &$  -85.69  $   &$  29.39$  &   $140.31$    &$  22  $\\
25  &   9064.3489   &$  -86.99  $   &$  18.71$  &   $168.46$    &$  11.21   $\\
26  &   9064.4351   &$  -   $   &$  -   $   &   $158.39$    &$  30.96   $\\
27  &   9064.4501   &$  -   $   &$  -   $   &   $143.44$    &$  11.02   $\\
28  &   9064.4654   &$  -   $   &$  -   $   &   $157.69$    &$  11.52   $\\
29  &   9064.4805   &$  -   $   &$  -   $   &   $137.11$    &$  15.42   $\\
30  &   9064.4956   &$  -78.27  $   &$  15.22$  &   $153.28$    &$  14.83   $\\
31  &   9065.3820   &$  97.01   $   &$  15.22$  &   $   -   $   &$  -   $\\
32  &   9065.3970   &$  86.97   $   &$  17.00$  &   $   -   $   &$  -   $\\
33  &   9065.4122   &$  82.29   $   &$  20.00$  &   $   -   $   &$  -   $\\
34  &   9065.5201   &$  -   $   &$  -   $   &   $-16.69$    &$  12.98   $\\
35  &   9065.5352   &$  -   $   &$  -   $   &   $-19.88$    &$  16.82   $\\
36  &   9068.4837   &$  -   $   &$  -   $   &   $-19.81$    &$  27.49   $\\
37  &   9068.4989   &$  -   $   &$  -   $   &   $   -9.48$  &$  23.13   $\\
38  &   9069.4358   &$  -   $   &$  -   $   &   $   14.23$  &$  17  $\\
39  &   9069.4510   &$  -   $   &$  -   $   &   $   15.88$  &$  16.7    $\\
40  &   9069.4665   &$  -   $   &$  -   $   &   $   1.42$   &$  15.9    $\\
41  &   9070.2203   &$  -   $   &$  -   $   &   $178.74$    &$  10.02   $\\
42  &   9127.3567   &$  13.44$  &$  15.23   $   &$  -   $   &$  -   $\\
43  &   9127.3720   &$  11.79$  &$  18.3    $   &$  -   $   &$  -   $\\
44  &   9129.3012   &$  -   $   &$  -   $   &   $   -6.91$  &$  15.37   $\\
45  &   9129.3171   &$  141.83  $   &$22.00 $   &   $-52.67$    &$  20.83   $\\
46  &   9129.3323   &$  -   $   &$  -   $   &   $-42.14$    &$  17  $\\
47  &   9304.5708   &$  -   $   &$  -   $   &   $-15.39$    &$  11.22   $\\
48  &   9304.5863   &$  -   $   &$  -   $   &   $-28.77$    &$  13  $\\
49  &   9304.6020   &$  -   $   &$  -   $   &   $-26.09$    &$  12.22   $\\
50  &   9304.6174   &$  -   $   &$  -   $   &   $-45.49$    &$  14.23   $\\
51  &   9308.5816   &$  16.73$  &$  15.23   $   &$  -   $   &$  -   $\\
52  &   9309.5552   &$  12.71   $   &$  16.23$  &   $   -   $   &$  -   $\\
53  &   9309.5715   &$  10.04   $   &$  14.00$  &   $   -   $   &$  -   $\\
54  &   9309.5870   &$  21.41   $   &$  17.21$  &   $   -   $   &$  -   $\\
55  &   9311.5704   &$  -15.39$ &$  17.22   $   &$  -   $   &$  -   $\\
56  &   9311.5853   &$  -16.06$ &$  16.23   $   &$  -   $   &$  -   $\\
57  &   9311.6004   &$  -17.39$ &$  16.22   $   &$  108.38  $   &$  15.22   $\\
\hline
\end{tabular}
\end{center}
\end{table}

\end{appendix}

\end{document}